\documentclass[aps,prd,twocolumn,showkeys,floats]{revtex4}
\usepackage{graphics,epsfig,color} 
\usepackage{graphicx}
\usepackage{amsmath,amssymb}

\newcommand\mordre[1] {O(c^{- {#1}})}
\begin{document}
\title{Explicit formulae for the two way time-transfer in the T2L2 experiment including the $J_2$ contribution to the Earth potential in a relativistic framework}

\author{O. Minazzoli*\footnote{email : olivier.minazzoli@oca.eu} ~, B. Chauvineau* ~, E. Samain\dag~, P. Exertier\dag~, P. Vrancken\dag~ and P. Guillemot \ddag}
\affiliation{* UNS, OCA-ARTEMIS UMR 6162, Observatoire de la Côte d'Azur, Avenue Copernic, 06130 Grasse, France
\\ \dag UNS, OCA-Geosciences Azur UMR 6526, Observatoire de la Côte d'Azur, Avenue Copernic, 06130 Grasse, France
\\ \ddag CNES, Avenue Edouard Belin, 31401 Toulouse cedex 9, France}

\begin{abstract}
The topic of this paper is to study the two way time-transfer problem between a
ground based station and a low orbit Earth's satellite, in the aim of an
application to the T2L2 experiment.\ The sudy is driven in a fully
relativistic framework.\ Because of the rapid increase in clock's
precision/measurements, the first term beyond the Earth's potential
monopolar term is explicitly taken into account. Explicit formulae, for both the distance and offset problems (definitions in the text) are proposed for the relevant applications.
\end{abstract}

\maketitle

\section{Introduction}

The idea of optical space-based time transfer had first been proposed to ESA in the seventies through the project which was named LASSO \cite{Fridelanceetal} (Laser Synchronization from stationary Orbit). Even though it was quite difficult to get some data with LASSO, the mission was a great success with some time transfers obtained within a time stability in the range of 100 ps. In 1994 OCA proposed to build a new generation of optical transfer based on some new technology: it was the beginning of the T2L2 project. As compared to LASSO, the objective was to improve the performances by at least 2 orders of magnitude.  After several proposals on the Mir Space Station, ISS, GIOVE, and Myriade, T2L2 was finally accepted in 2005 as a passenger instrument on the Altimetry Jason-2 satellite (\cite{Fridelanceetal2},\cite{Samainetal}).

The full exploitation of the T2L2 experiment requires a relativistic
description of the Earth gravitational field for both the definition of the
proper timescales relevant for the (ground-based and on board) clocks and the photons' trajectories involved by laser links.

The aim of the current paper is to propose explicit useful formulae dealing with
two-way experiments, like T2L2.\ We derive explicit formulae giving the ratio 
$\nu _{G}^{+}/\nu _{G}^{-}$ of the ground-based frequencies at reception and
emission respectively as a function of the required quantities (positions,
velocities,...) at a ground-based time. We also give
explicit expressions for the time dependance of (1) the ground proper time
needed for a photon to go to the satellite and to come back and (2) the offset between the on board and the ground based
clocks. All the quantities are explicitely written as a function of the time attached to a unique event (emission or reception). To anticipate the performance increase of future
clocks, these formulae take the $J_{2}$ term in the Earth
gravitational field effect on the photons' orbits into account.

Blanchet et al \cite{bstw01} derive explicit formulae in both the one-way and the
two-way problems.\ However, in the two-way case, the ratio $\nu _{G}^{+}/\nu
_{G}^{-}$ of the ground-based frequencies at reception and emission
respectively and other useful expressions are not explicitly given as a function of the required quantities at a unique time.\
The current paper proposes such formulae, which can be directly used in
practical applications.\ Besides, the photon's orbit is determined in \cite{bstw01}
using the monopolar Earth gravitational term only. (A more complete
gravitational modelization is used in \cite{bstw01}, but for the link between proper
and coordinate timescales only.)\ In the present paper, we include the $J_{2}
$ gravitational contribution to light propagation, which will be useful as
soon as the $10^{-14}\;s$ level of precision is required, as it should be in the future ACES experiment \cite{acespubli}, for instance. Using the
Synge world function approach \cite{Synge}, Linet and
Teyssandier \cite{lt02} have obtained explicit expressions of laser link in the
one-way case for general axisymetric gravitational field, but no formulae in
the two-way case is given. This problem has been reconsidered by Le
Poncin-Lafitte et al \cite{llt04}, but they give explicit formulae in the spherical case and for the one-way
problem. Let us remark that, from a theoretical point of view, the applicability of
the Synge world function approach can be limited to regions of spatial
extension of the order of $2R_{Sun}(R_{Sun}/R_{Schw})\sim
2000\;A.U.\sim 0.01\;pc$ in the worse case (light rays grazzing the Sun and
considering points at typically $\sim 1000\;A.U.$). The last remark is
obviously not relevant for solar system applications, but is relevant for interstellar scales.

\section{Experimental Description}
\subsection{Principe}
The experiment principle is issued from satellite laser ranging techniques. T2L2 permits to synchronize remote ground clocks and compare their frequency stabilities with a performance never reached before. T2L2 allows synchronization of a ground and space clock and measurement of the stability of remote ground clocks over continental distances, itself having a time stability in the range of 1 ps over 1000 s.

\begin{figure}
\begin{center}
\includegraphics[scale=0.2]{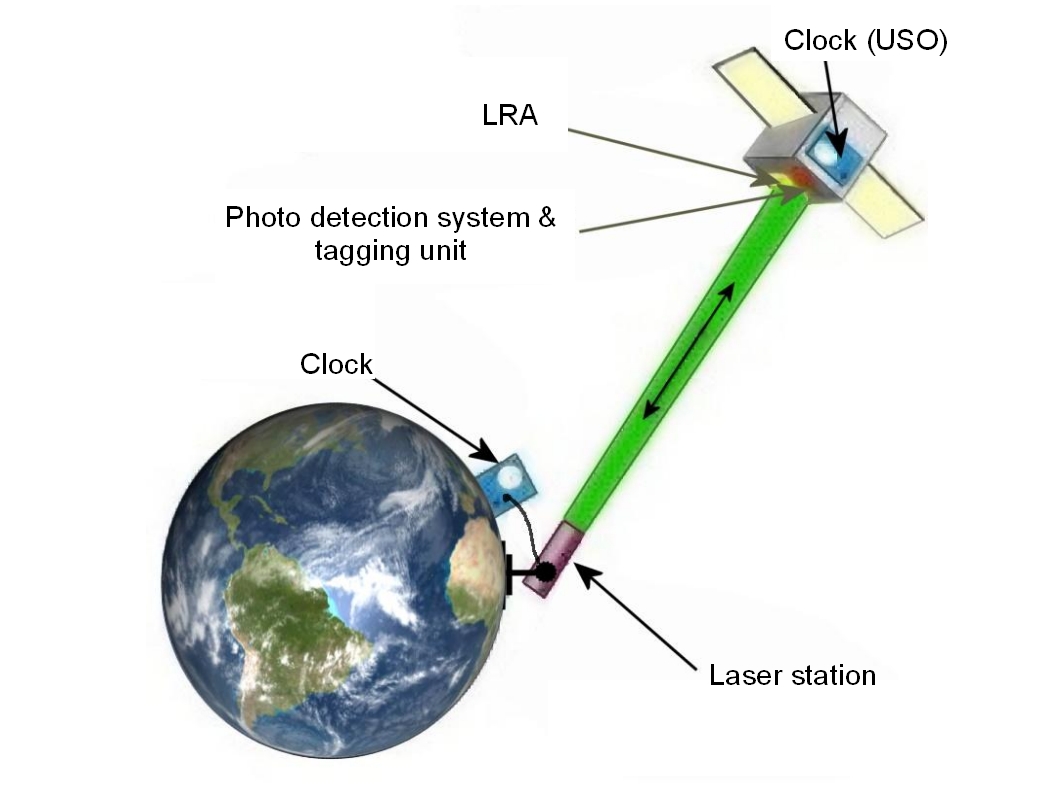}
\caption{Ground to space time transfer with T2L2}
\label{fig:fig_et1}
\end{center}
\end{figure}

\begin{figure}
\begin{center}
\includegraphics[scale=0.1]{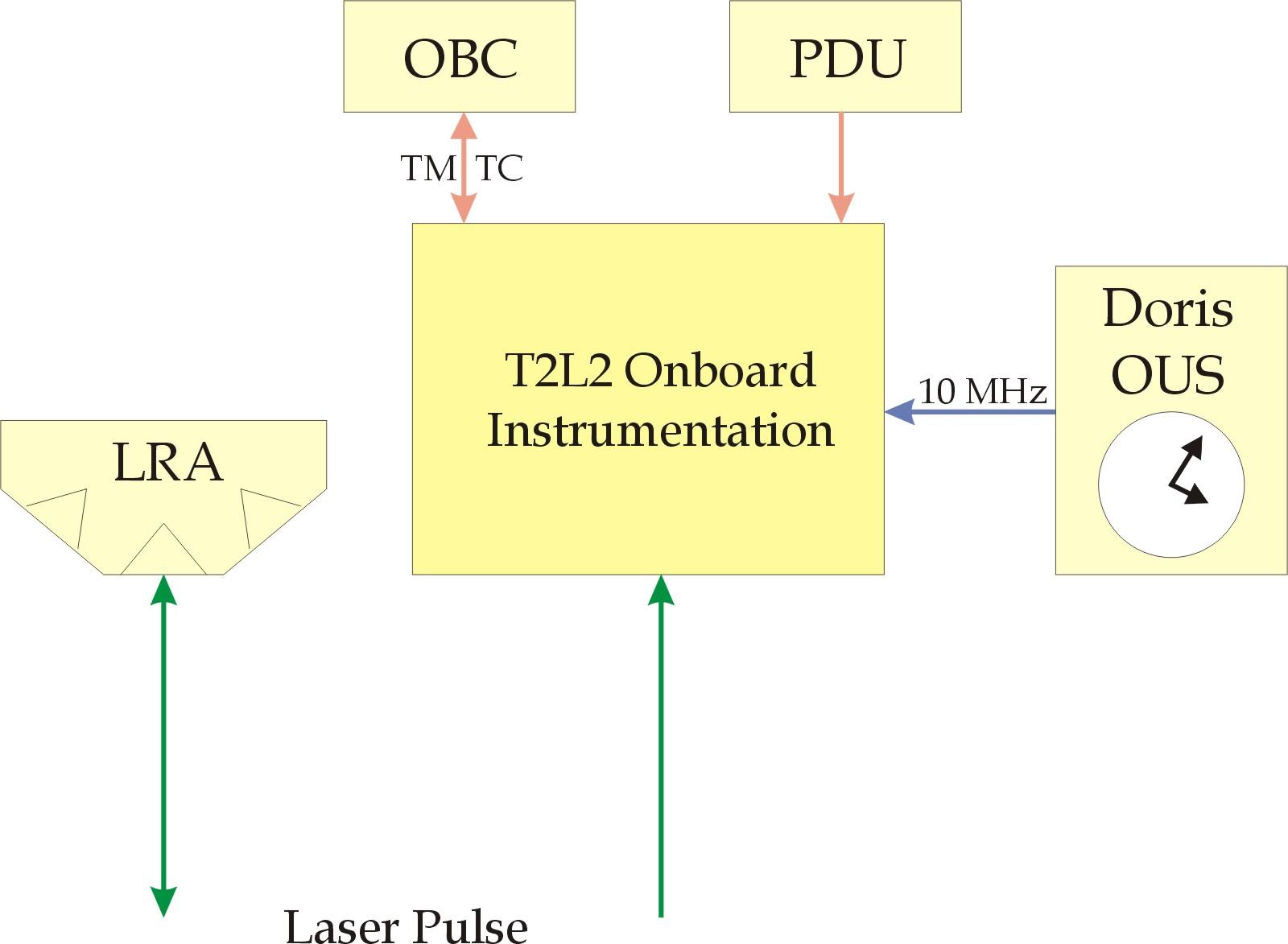}
\caption{Global synoptic of the T2L2 instrument on Jason-2 together with the LRA and the DORIS USO}
\label{fig:fig_et2}
\end{center}
\end{figure}

The principle is based on light pulses propagation between ground laser stations and a satellite equiped with a photo detection system and a time-tagging unit (fig. \ref{fig:fig_et1}). The space instrument uses a Laser Ranging Array (LRA) which is also used for some laser ranging purposes and the Doris ultra-stable oscillator (USO) fundamentally used by the DORIS positioning system (fig. \ref{fig:fig_et2}). The USO is linked to the time tagging unit.  It is the onboard reference clock of the project. 

A laser station at ground emits short asynchronous laser pulses towards the satellite. The LRA returns a fraction of the received photons back to the station. To describe roughly the main idea of the principle of the measurements, the station records the start ($t_\textrm{start}$) and return ($t_\textrm{return}$) times of each light pulse. The T2L2 payload records the arrival time ($t_\textrm{board}$) in the temporal reference frame of the onboard oscillator. These data are regularly downloaded every 2 hours to the ground via a regular microwave communication link.

For a given light pulse emitted from ground, the time offset $\theta_\textrm{GS}$ between ground clock G and space clock S is deduced from the measurement triplets {$t_\textrm{start}$, $t_\textrm{board}$, $t_\textrm{return}$} with the following time equation:
\begin{equation}
\label{eq:offset_newton}
\theta_\textrm{GS} = (t_\textrm{start} + t_\textrm{return})/2 - t_\textrm{board} + \epsilon,
\end{equation}
where $\epsilon$ is a corrective term, including atmospheric effects for instance. The time transfer between several laser stations at ground (A, B, C, ...) is then deduced from the difference between each ground to space time transfer.

However, a formulae like (\ref{eq:offset_newton}) is based on some inadequate physical concepts
from the start. Obviously, it involves only a rough notion of time, leaving
in the dark the fact that different proper times have to be considered in
the problem, as it is well known from the relativistic gravity theory. In
fact, (\ref{eq:offset_newton}) is correct as long as the newtonian description of (space-)time
correctly describes the physics. However, at the level of precision reached
by experiments from years ago, it is well known it is no longer the case.\
Of course, it is possible to put the so-called ''relativistic corrections''
into the $\epsilon $ corrective term in (\ref{eq:offset_newton}).\ However, it is better to
reformulate the problem directly in the relativistic framework, this way
ensuring that all the ''relativistic effects'' will be consistently taken
into account at the considered order.

Depending on the distance between the laser ground stations (roughly 6000 km for Jason2) T2L2 can be operated in a common view mode or in a non-common view mode. In common view configuration, with two laser ranging stations A and B firing towards the satellite simultaneously, the noise of the on-board oscillator has to be considered over time interval equal to the delay between consecutive laser pulses. In such a way, the noise coming from the onboard oscillator can be considered negligible in the global error budget.  In a non-common view mode, the temporal information is carried by the satellite local oscillator over the distance separating the two ground stations visibility and the USO's noise becomes an important part. 

\subsection{T2L2 on JASON2}
Jason-2 is a French-American follow-on mission to Jason-1 and Topex/Poseidon. Conducted by NASA and CNES, its goal is to study the internal structure and dynamics of ocean currents mainly by radar altimetry. For the needs of precise determination of the satellite orbit, three independent positioning systems are also included: a Doris transponder, a GPS receiver and the LRA (Laser Ranging Array) retroreflector. The satellite was placed in a 1,336 km orbit with $66^o$ inclination by a Delta launcher. The time interval between two passes varies from 2 to 14 hours with an average duration of about 1000 s.
\begin{figure}
\begin{center}
\includegraphics[scale=0.1]{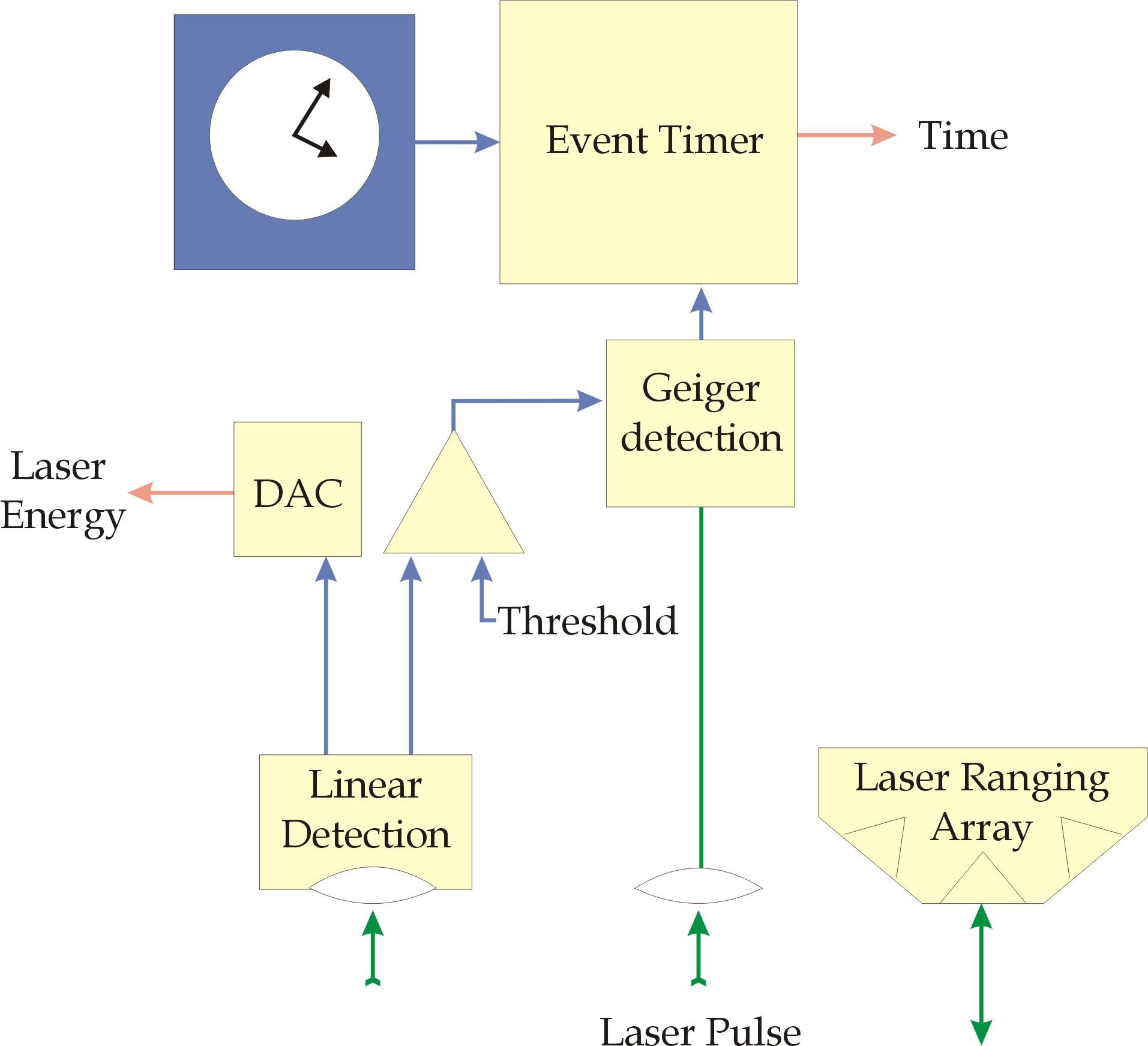}
\caption{Synoptic of the T2L2 space instrument}
\label{fig:fig_et3}
\end{center}
\end{figure}
The T2L2 instrumentation contains two main subsystems: of a photo detection module and an event timer (fig. \ref{fig:fig_et3}). The photo detection unit is made with 2 avalanche photo diodes. One is working in a special non-linear "Geiger" mode for precise chronometry. The other works in linear gain mode in order to trigger the whole detection chain and to measure the received optical energy and the continuous noise flux originating in the earth albedo. To minimize the false detection rate, the detection threshold may be adjusted either by remote control or automatically as a function of the continuous flux measurement. Some elements of the photo detection unit  are located outside the main Jason-2 payload on a boom which support also the LRA. These elements are oriented toward the earth surface. The rest of the detection is, together with event timer,  inside an unique package located inside the Jason-2 payload. All this equipment  has a mass of 10 kg and a power consumption of 42 W.

\subsection{T2L2 Objectives}
The objectives of the T2L2 experiment on Jason-2 are threefold:
\begin{itemize}
\item Validation of optical time transfer, including the validation of the experiment, its time stability and accuracy. It should further allow to demonstrate one-way laser ranging
\item Scientific applications concerning time and frequency metrology allowing the calibration of radiofrequency time transfer (GPS and Two-Way), fundamental physics with the measurement of light speed anisotropy and alpha fine structure constant, Earth observation and very long baseline interferometry (VLBI).
\item Characterization of the on-board Doris oscillator, especially above the South Atlantic Anomaly (SAA). 
\end{itemize}
\subsection{First results}

T2L2 relies on the international laser ranging network that permanently contributes to 
the International Earth Reference System (terrestrial frame and Earth's rotation), and to
the traking of several scientific space missions. Since 1998, this network has been 
organised as an International Service (named ILRS), as it was the case for the GPS 
and VLBI geodetic space techniques. 
Among 35 Satellite Laser Ranging (SLR) stations, 17 of them provide full rate 
ranging data on Jason-2 that are downloaded to the European Data Center 
(EDC in Muechen, Germany) regularly; the data files are uploaded by the T2L2 
Scientific Mission Center, located in Grasse, France. 
Among these stations, 10 of them use the new data format that permits to 
extract the start time of laser pulses at the ps level, whereas the others use the
current format. This data format was fitted for laser ranging only and use a 
maximum of 12 digits to record the start epoch of the laser pulses (corresponding 
to 100 ns). 

Several steps are necessary to proceed together ground SLR and on board T2L2 
data sets (the T2L2 data files are uploaded from CNES, Toulouse to the Mission 
Center every day with a time delay of 1-2 days):
\begin{itemize}
\item selection of data relatively to the SLR stations and satellite passes, 
\item extraction of measurement triplets from both data sets, 
\item estimation of instrumental corrections, 
\item determination of the time of light travel between the 
SLR station and the T2L2 space instrument, 
\item computation of the ground to space time transfer. 
\end{itemize}

\subsubsection{Data selection}

The first step of the data treatment consists in determining
the absolute frequency offset and the phase delay between
the space clock and the UTC time scale, along the time. 
The goal is to compute the on board date of each detected 
optical event (i.e., arrival time of emitted laser pulses) in a time 
scale that is identical to UTC at the picoseconde level. Because the second step of the data 
treatment (Triplets extraction), is based on a direct 
comparison of the dates of T2L2 data and ground SLR data. 

The T2L2 space instrument being connected to the GPS of the 
Jason-2 plateform, we use Pulse Per Second (PPS) that are 
permanently emitted by this system at $\pm1 \mu$sec and 
then recorded by T2L2 in the space clock time scale. That
permits to establish a phase link and an absolute frequency 
offset between the space clock and the UTC-GPS time scale, 
respectively to 0.3 $\mu$sec and to 2-3.$10^{-11}$ Hz (over 
5000 sec). 

\subsubsection{Triplets extraction}

The second step of the data treatment consists in recognizing 
from all the events recorded by T2L2 those corresponding to emitted 
laser pulses. Because the events coming from the SLR stations are 
blended together, it is necessary to find from which station a given 
event is coming up. On the other hand, due to the combination of 
several parameters such as (i) the low energy level used by current 
SLR systems (around 50-100 mJ and even lower), (ii) the energy 
dispersion due to atmospheric effects, (iii) and the aperture size of the 
T2L2 optical module relatively to the diameter of the laser beam, 
all the emitted laser pulses don't have a corresponding on board 
event. 
As a statistical result, T2L2 is detecting more events from SLR stations 
that use a higher energy level for their laser pulses (there is a 
factor 2 to 3 between the stations). 

\subsubsection{Instrumental corrections}

Several instrumental corrections must be estimated both for 
the ground and space segments. At ground, the accuracy of the 
range is obtained by a calibration process which uses a calibration
target located at a known position by a classical survey at the mm level. 
At the satellite, several parameters have to be taken into account as 
(i) the geometrical delay between the reference points of the LRA and 
the T2L2 optical module, (ii) the time walk compensation of the photo
detection module which is sensitive to the number of received
photons, (iii) the event timer calibration based on an internal 
measurement process. 

All these corrections are computed at the ps level, using
the available mechanical, optical and electronical measurements 
that were determined for all Jason-2 scientific equipments before
its launch. In addition, the stellar sensors of the satellite are providing
the attitude informations needed to compute the geometrical effects. 

\subsubsection{Ground to space time transfer first results}

The ground to space time transfers that have been selected use 
a set of SLR data from very performant ground stations that, in addition, 
are equipped by a Hydrogen Maser or a Cesium clock as a permanent 
system of time. Thus, the present analysis takes into account data sets
from Wettzell (Germany), Matera (Italy), Changchun (China), and Grasse 
(France) stations. 

Nevertheless, if all these selected SLR stations provide data with an 
optimal precision (range and start times), a noise is introduced in the 
time transfer by the time intervals measured at ground. It mainly comes 
from the quality of the laser beam (i.e., the pulse width), the ground 
detection module (a photo detector), the optical behaviour of the on 
board LRA, and the atmosphere effects (two times); the current 
budget error is of 20 to 35 ps depending of each station. 
By taking into account the cinematics between the satellite and the 
ground station during a satellite pass, it is possible to compute a 
synthetic time of flight by averaging the row data locally over an 
integration duration of a few tens of seconds. The noise is thus 
decreased by a factor $1 / \sqrt N$ where $N$ is the number of row 
data. If that computation is done properly from a data set to another
it permits also to reduce the slight but significant differences between 
the laser stations. 

The noise of the time of light travel being reduced to 4-5 ps inside the 
time transfer, the time stability can be represented by a computed time 
variance between roughly 1 s and 600 s. As an example, the time stability 
of the Wettzell's Hydrogen Maser compared to the T2L2's DORIS quartz
oscillator is 40 ps at 1 s and 7 ps at 30 s. For time integration greater 
than 30 s this measurement is limited by the DORIS time stability which 
is 5 ps at 30 s and 10 ps at 100 s. 

\subsubsection{Improvements}

Now, we can routinely compute time transfers between different ground 
clocks and the space DORIS oscillator. Nevertheless, some work is still
required to improve the instrumental model (physical behaviour of the 
hardware) to the level of 1-2 ps at 30 s. 
As examples of this model, (1) the T2L2 photo detection module is sensitive 
to the number of incoming photons that should be properly corrected, and
(2) there is still a data filtering to develop in order to avoid Solar events that 
are detected by T2L2 just 1-3 ns before the real pulse event. 

But the current estimation of the T2L2 performances seem to be in a good 
accordance with the previous specifications of the overall project thus
justifying the present relativistics equations to be applied. 


\section{The space-time model used for time transfer}
\label{sec:model}

In GR, photons follow null geodesics of the space-time metric $g_{\mu \nu }$%
. We write the metric under the form 
\begin{equation}
ds^{2}=g_{00}c^{2}dt^{2}+2g_{0i}cdtdx^{i}+g_{ij}dx^{i}dx^{j}  \label{iv-1},
\end{equation}
where $ct=x^0$. In the IAU2000 recommendations, terms in $c^{-4}$ are neglected in $g_{ij}$,
while they are taken into account in both $g_{00}$ and $g_{0i}$ (with no
contribution in $g_{0i}$)\ \cite{skp03}.\ Since we are interested in light
motion, $dx\sim c dt$, considering $%
c^{-4}$ terms in $g_{00}$ (and $g_{0i}$) would be meaningless.\ Hence, the metric to be considered writes
\begin{equation}
ds^{2}=\left( -1+\frac{2w}{c^{2}}\right) c^{2}dt^{2}-\frac{8w^{i}}{c^{3}}%
cdtdx^{i}+\left( 1+\frac{2w}{c^{2}}\right) \delta _{ij}dx^{i}dx^{j}
\label{iv-2}
\end{equation}
where \cite{skp03} 
\begin{eqnarray}
w\left( t,x^{k}\right) &=&G\int \frac{\sigma \left( t,y^{k}\right) }{\left\|
y-x\right\| }d^{3}y  \label{iv-3} \\
w^{i}\left( t,x^{k}\right) &=&G\int \frac{\sigma ^{i}\left( t,y^{k}\right) }{%
\left\| y-x\right\| }d^{3}y  \notag
\end{eqnarray}
$\sigma $\ and $\sigma ^{i}$ being the gravitational mass and
mass current respectively.\ $\left\| u\right\| $ is a short notation for $%
\sqrt{u^{k}u^{k}}$. Hence, $w$ corresponds to the potential in Newtonian gravity.\ Let us write $w$ under the form 
\begin{eqnarray}
w &=&w_{E}+w_{ext}  \label{iv-4} \\
w_{E} &=&w_{E,m}+w_{E,J_{2}}+w_{E,up}  \notag
\end{eqnarray}
where $w_{E}$ is the Earth potential, written as the sum of the monopolar
term $w_{E,m}$, the oblateness term $w_{E,J_{2}}$, and the other terms $%
w_{E,up}$, when developed in spherical harmonics, and $w_{ext}$\ the
external potentials, caused by the presence of the Moon and the Sun
essentially. The orders of magnitude of these different terms are given by 
\begin{eqnarray}
\frac{w_{E,m}}{c^{2}} &\sim &\frac{GM_{E}}{rc^{2}}\sim 10^{-9}  \notag
\\
\frac{w_{E,J_{2}}}{c^{2}} &\sim &J_{2}\frac{GM_{E}}{rc^{2}}\frac{R_{E}^{2}}{%
r^{2}}\sim 10^{-3}w_{E,m}\sim 10^{-12}  \notag \\
\frac{w_{E,up}}{c^{2}} &\sim &10^{-2}w_{E,J_{2}}\sim 10^{-14}  \notag \\
\frac{w_{ext}}{c^{2}} &\sim &x^{i}x^{j}\partial _{i}\partial _{j}U\sim \frac{%
GM}{Lc^{2}}\frac{r^{2}}{L^{2}}=\frac{GM_{E}}{rc^{2}}\frac{M}{M_{E}}\left( 
\frac{r}{L}\right) ^{3}  \notag
\end{eqnarray}
where $M_{E}$\ and $R_{E}$\ are Earth's mass and radius, and $U$ the
Newtonian potential for a remote body of mass $M$, at distance $%
L$. It results that, for satellites at low altitude (say $\lesssim 1000\;km$%
), the term $w_{E,m}$\ induces corrections which can reach $\sim 10^{-11}\;s$
for the time transfer. The effect of the $J_{2}$\ term can then be of the
order of $10^{-14}\;s$. The following terms in the Earth potential
development results into corrections of order $10^{-16}\;s$.\ For both the
Sun and the Moon, $w_{ext}$ is of order $10^{-19}$.

The time-space term in the metric (dragging effect) is of order 
\begin{equation}
\frac{8w^{i}}{c^{3}}\sim \frac{8G}{c^{3}}\int \frac{\rho V^{i}\left(
t,y^{k}\right) }{\left\| y-x\right\| }d^{3}y  \label{iv-6}
\end{equation}
where $\rho $\ is the Earth's density, and $V$\ the field of matter velocity
inside the Earth. Assuming a spherical rigidly rotating Earth, with uniform
density and angular velocity $\omega $, the expression (\ref{iv-6}) leads
to, for a satellite located in the equatorial plane at altitude $h$ 
\begin{equation}
\frac{8\left\| w^{i}\right\| }{c^{3}}=\frac{8}{5}\frac{GM_{E}}{R_{E}c^{2}}%
\frac{\omega R_{E}}{c}\left( \frac{R_{E}}{R_{E}+h}\right) ^{2}\sim 2.10^{-15}
\label{iv-7}
\end{equation}
for a low altitude satellite. The resulting correction to the photon flying
time is then expected to be of the order of $10^{-17}\;s$.

In a foreseeable future, time transfer precision of order $10^{-15} s$ is expected. Hence, for low altitude satelites, one has to consider the $J_2$ term in Earth potential while all the other terms may be discarded.

\section{Application}
\label{sec:appl}
\subsection{The photon orbit}

From the previous section, we will consider in this section the time
transfer problem in a space-time described by the metric 
\begin{equation}
ds^{2}=\left( -1+\frac{2w}{c^{2}}\right) c^{2}dt^{2}+\left( 1+\gamma \frac{2w%
}{c^{2}}\right) \delta _{ij}dx^{i}dx^{j}  \label{v-1}
\end{equation}
where $\gamma $ is a PPN parameter including
possible extensions of GR, like scalar-tensor (ST) theories \cite{w06}. The
potential $w$ is given by 
\begin{equation}
w=\overset{-}{w}+J_{2}\overset{\sim }{w}  \label{v-2}
\end{equation}
with 
\begin{equation}
\overset{-}{w}=\frac{GM_{E}}{r}\text{ \ \ \ \ \ and \ \ \ \ \ }\overset{\sim 
}{w}=\frac{GM_{E} R^2}{2r^{3}}\left( 1-3\frac{z^{2}}{r^{2}}\right)  \label{v-3}
\end{equation}
$R$ being a parameter of the order of the Earth radius. Since we are
interested in photon propagation, one has to solve the geodesic equation 
\begin{equation}
\frac{d}{d\lambda }\left( g_{\alpha \beta }k^{\beta }\right) =\frac{1}{2}%
k^{\mu }k^{\nu }\partial _{\alpha }g_{\mu \nu }\text{ \ \ \ \ \ with \ \ \ \
\ }k^{\alpha }=\frac{dx^{\alpha }}{d\lambda }  \label{v-4}
\end{equation}
($\lambda $ being an affine parameter). The wave quadri-vector $k$
satisfies the null condition 
\begin{equation}
g_{\alpha \beta }k^{\alpha }k^{\beta }=0  \label{v-5}
\end{equation}
expressing the motion occurs on light cones. The way chosen to solve
equations (\ref{v-4}) and (\ref{v-5}), up to first order in $w$, is the way
chosen in \cite{crvp05}, in which only the $\overset{-}{w}$-term was considered
in the potential. The aim is to derive the trajectory of the photon $%
x^{i}\left( x^{0}\right) $, obtained after the elimination of the affine
parameter. (This elimination can be made both after the integration of the
equations or at the level of the differential equation (\ref{v-4}) itself.)

Up to $\mordre{3}$ terms, the result reads
\begin{equation}
x_{ph}^{i}\left( t,n^{k},x_{A}^{k}\left( t_{e}\right) \right) =x_{0}^{i}+n^{i}ct + \overset{\left( 1\right) }{x}%
_{ph}^{i}\left( t,n^{k},x_{A}^{k}\left( t_{e}\right) \right) \label{v-6} 
\end{equation}
where, using the short notation $m=GM_{E}/c^{2}$%
\begin{eqnarray}
\overset{\left( 1\right) }{x}%
_{ph}^{i}\left( t,n^{k},x_{A}^{k}\left( t_{e}\right) \right)&=&\frac{1+\gamma }{2}mn^{i}\ln \frac{r-\Lambda }{r+\Lambda }-\left(
1+\gamma \right) m\frac{\xi ^{i}}{K^{2}}r  \notag \\
&&+\left( 1+\gamma \right) J_{2}mR^{2}\Psi ^{i}\left( t\right)  +Z^i \notag
\end{eqnarray}
with 
\begin{eqnarray}
&\Psi& ^{i}\left( t\right) =\left[ -\frac{\xi ^{i}+2\delta _{3i}\xi ^{3}}{%
K^{4}}+\frac{n^{3}n^{3}\xi ^{i}+2n^{3}n^{i}\xi ^{3}}{K^{4}}+4\frac{\xi
^{3}\xi ^{3}\xi ^{i}}{K^{6}}\right] r  \label{v-7} \notag \\
&&+\left[ \frac{-n^{i}+2\delta _{3i}n^{3}}{2K^{2}}-\frac{n^{3}n^{3}n^{i}}{%
2K^{2}}+\frac{n^{i}\xi ^{3}\xi ^{3}-2n^{3}\xi ^{3}\xi ^{i}}{K^{4}}\right] 
\frac{\Lambda }{r}  \notag \\
&&+\left[ \frac{\xi ^{i}+2\delta _{3i}\xi ^{3}}{2K^{2}}-\frac{n^{3}n^{3}\xi
^{i}+n^{3}n^{i}\xi ^{3}}{K^{2}}-2\frac{\xi ^{3}\xi ^{3}\xi ^{i}}{K^{4}}%
\right] \frac{1}{r}  \notag \\
&&+\left[ -\frac{1}{2}n^{3}n^{3}n^{i}+\frac{n^{i}\xi ^{3}\xi ^{3}-2n^{3}\xi
^{3}\xi ^{i}}{2K^{2}}\right] \frac{\Lambda }{r^{3}}  \notag \\
&&+\left[ -n^{3}n^{i}\xi ^{3}+\frac{1}{2}n^{3}n^{3}\xi ^{i}-\frac{\xi
^{3}\xi ^{3}\xi ^{i}}{2K^{2}}\right] \frac{1}{r^{3}}  \end{eqnarray}
where $x_{0}^{i}$ is the position at $t=0$ (the integration constants $Z^{i}$%
\ being adjusted this way \cite{sf2a2007}). The three numbers $n^{i}$ are also integration
constants satisfying the normalization condition 
\begin{equation}
n^{i}n^{i}=1.  \label{v-8}
\end{equation}
In the first order terms, one has 
\begin{equation}
\Lambda =ct+n^{i}x_{0}^{i}  \label{v-9}
\end{equation}
\begin{equation}
K^{2}=x_{0}^{i}x_{0}^{i}-\left( n^{i}x_{0}^{i}\right) ^{2}  \label{v-10}
\end{equation}
\begin{equation}
r=\sqrt{K^{2}+\Lambda ^{2}}  \label{v-11}
\end{equation}
\begin{equation}
\xi ^{i}=x_{0}^{i}-n^{i}n^{k}x_{0}^{k}.  \label{v-12}
\end{equation}
Let us remark that $\xi^{i}\xi ^{i}=K^{2}$ and $n^{i}\xi ^{i}=0$.
\subsection{Time transfer and useful formulae}
\subsubsection{The time transfer}

Let us now consider two photons emitted from the station A and received by
the station B.\ Let $d\tau _{A}$ and $d\tau _{B}$ be the proper time
intervals between, respectively, the two emissions and the two receptions, and $dt_{A}$, $%
dt_{B}$ the corresponding coordinate time intervals.\ Let $x_{A0}^{i}$, $x_{B0}^{i}$ be the positions and $v_{A0}^{i}$, $v_{B0}^{i}$ the coordinate velocities of A and B at
the coordinate time corresponding to the emission of the first photon by A.

Since both A and B have velocities corresponding to orbital free motions
(for a satellite) or smaller (for a ground station), the velocities of both
are at best of the order of $\sqrt{GM_{E}/r}$, in such a way that the $g_{0i}
$ term in (\ref{iv-2}) contributes as a  $c^{-4}$ term in
each station proper time. The ground station's potential has to be modelized more precisely than in (\ref{v-2}) \cite{bstw01}. Hence we generally denote it by $W$ and not by $w$ which is its approximation (\ref{v-2}) ($w$ is only suitable for a satellite and one can replace $W$ by $w$ in this case in practical calculations). One has
\begin{equation}
d\tau _{A}=dt_{A}\left[ 1-\frac{W_{A}}{c^{2}}-\frac{v_{A}^{2}}{2c^{2}}%
+\mordre{4} \right]   \label{v-13}
\end{equation}
and an analogous expression for $d\tau _{B}$. Since $t_{B}=t_{A}+t$, where $t$%
\ is the time transfer (to be determined later),\ this leads to 
\begin{eqnarray}
\frac{d\tau _{B}}{d\tau _{A}}=&&\left[ 1+\left( \frac{dt}{dt_{A}}\right) _{0}%
\right] \times \notag \\&& \left[ 1+\frac{W_{A0}-W_{B}}{c^{2}}+\frac{v_{A0}^{2}-v_{B}^{2}}{%
2c^{2}}+\mordre{4} \right] .  \label{v-14}
\end{eqnarray}
The indexes $0$ recall that the corresponding quantities have to be computed
at the first photon emission time. The function $t\left( t_{e}\right) $ as a
function of the emission time $t_{e}$ ($=t_A$) is implicitely obtained from the
equation defining the interception of the photon by the station B 
\begin{equation}
x_{ph}^{i}\left( t,n^{k},x_{A}^{k}\left( t_{e}\right) \right)
=x_{B}^{i}\left( t,x_{B}^{k}\left( t_{e}\right) ,v_{B}^{k}\left(
t_{e}\right) ,...\right)   \label{v-15}
\end{equation}
where one has explicitely written the dependence of the station B in initial (i.e. at the emission time) orbital parameters. The function $x_{ph}^{i}\left(
t,n^{k},x_{A}^{k}\left( t_{e}\right) \right) $ is explicitely given by (\ref
{v-6}), (\ref{v-7}) up to first order. The function $x_{B}^{i}\left(
t,x_{B}^{k}\left( t_{e}\right) ,v_{B}^{k}\left( t_{e}\right) ,...\right) $
can be obtained from the geodesic equation if the station B is a satellite while it is
given by an Earth rotation model in the case of a ground based station.\
Let us write these functions as 
\begin{eqnarray}
x_{ph}^{i}\left( t,n^{k},x_{A}^{k}\left( t_{e}\right) \right)
&=&x_{A}^{i}\left( t_{e}\right) +n^{i}ct \notag \\&& +\overset{\left( 1\right) }{x}%
_{ph}^{i}\left( t,n^{k},x_{A}^{k}\left( t_{e}\right) \right) \notag \\&& + \overset{%
\left( 3/2\right) }{x}_{ph}^{i}\left( t,n^{k},x_{A}^{k}\left( t_{e}\right)
\right)   \label{v-16}
\end{eqnarray}
and 
\begin{eqnarray}
\label{v-17}
x_{B}^{i}&&\left( t,x_{B}^{k}\left( t_{e}\right) ,v_{B}^{k}\left( t_{e}\right)
,...\right) = \\&& x_{B}^{i}\left( t_{e}\right) +tv_{B}^{i}\left( t_{e}\right) \notag  +%
\frac{1}{2}t^{2}a_{B}^{i}\left( t_{e}\right) +\frac{1}{6}t^{3}b_{B}^{i}%
\left( t_{e}\right)   
\end{eqnarray}
where $\overset{\left(
3/2\right) }{x}_{ph}^{i}$ corresponds to the (unknown) $3/2$ order part of
the photon orbit (contribution of the $g_{0i}$ term). The coefficients $%
a_{B}^{i}\left( t_{e}\right) $ and $b_{B}^{i}\left( t_{e}\right) $\
correspond to the acceleration of the station B and its derivative at the
emission time. The four r.h.s. terms in (\ref{v-17}) are respectively
zeroth, first, second and third order terms in $c^{-1}$ (since $%
tv_{B}=ct\left( v_B/c\right) $, $t^{2}a_{B}=\left( ct\right) ^{2}\left(
a_{B}/c^{2}\right) $ and so on).\\
Solving (\ref{v-17}) leads to 
\begin{eqnarray}
ct &=&\epsilon \overset{\left( 0\right) }{\theta }+\overset{\left(
1/2\right) }{\theta }+\epsilon \overset{\left( 1\right) }{\theta }+\mordre{3}   \label{v-36} \\
n^{i} &=&\epsilon \overset{\left( 0\right) }{\eta }^{i}+\overset{\left(
1/2\right) }{\eta }^{i}+\epsilon \overset{\left( 1\right) }{\eta }%
^{i}+\mordre{3}  \notag
\end{eqnarray}
where $\epsilon = \pm $ and with (quantities at time $t_{e}$)
\begin{eqnarray*}
\overset{\left( 0\right) }{\theta } &=&d\text{ \ \  
\ where \ \ }d\equiv\sqrt{\left( x_{B}^{i}-x_{A}^{i}\right) \left(
x_{B}^{i}-x_{A}^{i}\right) } \\
\overset{\left( 0\right) }{\eta }^{i} &=&\frac{x_{B}^{i}-x_{A}^{i}}{d} \\
\overset{\left( 1/2\right) }{\theta } &=&d\overset{\left( 0\right) }{\eta }%
^{i}\frac{v_{B}^{i}}{c} \\
\overset{\left( 1/2\right) }{\eta }^{i} &=&p^{ik}\frac{v_{B}^{k}}{c}\text{ \
\ \ \ \ \ \ \ \ where \ \ }p^{ik}=\delta ^{ik}-\overset{\left( 0\right) }{%
\eta }^{i}\overset{\left( 0\right) }{\eta }^{k} \\
\overset{\left( 1\right) }{\theta } &=&\frac{1}{2}d\left( \delta ^{kl}+%
\overset{\left( 0\right) }{\eta }^{k}\overset{\left( 0\right) }{\eta }%
^{l}\right) \frac{v_{B}^{k}v_{B}^{l}}{c^{2}}+\frac{1}{2}d^{2}\overset{\left(
0\right) }{\eta }^{k}\frac{a_{B}^{k}}{c^{2}}\\
&&-\overset{\left( 0\right) }{\eta 
}^{k}\overset{\left( 1\right) }{x}_{ph}^{k}\left( \frac{d}{c},\overset{%
\left( 0\right) }{\eta }^{m},x_{A}^{m}\right)  \\
\overset{\left( 1\right) }{\eta }^{i} &=&p^{ik}\left[ \frac{1}{2}d\frac{%
a_{B}^{k}}{c^{2}}-\frac{1}{d}\overset{\left( 1\right) }{x}_{ph}^{k}\left( 
\frac{d}{c},\overset{\left( 0\right) }{\eta }^{m},x_{A}^{m}\right) \right]\notag \\
&-&\frac{1}{2}\overset{\left( 0\right) }{\eta }^{i}p^{kl}\frac{%
v_{B}^{k}v_{B}^{l}}{c^{2}}
\end{eqnarray*}
$\epsilon=+$ corresponds to a photon travelling from A to B ($t_A<t_B$) while $\epsilon=-$ corresponds to a photon travelling from B to A ($t_A>t_B$).
\\
Considering now two emissions, at times $t_{e}$
and $t_{e}+dt_{e}$, the transfer times are $t$ and $t+dt$, while the
''directions'' are $n^{i}$and $n^{i}+dn^{i}$. Differentiating $%
t^{3}b_{B}^{i}\left( t_{e}\right) $ with respect to $t_{e}$ leads to a $%
c^{-4}$ term, since B has at best a satellite like velocity. Differentiating
(\ref{v-15}), (\ref{v-16}), (\ref{v-17}) and retaining only terms up to $%
c^{-3}$ leads to 
\begin{eqnarray}
\left[ cn^{i}-v_{B}^{i}-ta_{B}^{i}+\frac{\partial \overset{\left( 1\right) }{%
x}_{ph}^{i}}{\partial t}-\frac{1}{2}t^{2}b_{B}^{i}+\frac{\partial \overset{%
\left( 3/2\right) }{x}_{ph}^{i}}{\partial t}\right] dt   \notag \\
+\left[ v_{A}^{i}-v_{B}^{i}-ta_{B}^{i}-\frac{1}{2}t^{2}b_{B}^{i}+v_{A}^{k}%
\frac{\partial \overset{\left( 1\right) }{x}_{ph}^{i}}{\partial x_{A}^{k}}%
\right] dt_{e}  \notag \\
+\left[ ct\delta ^{ik}+\frac{\partial \overset{\left( 1\right) }{x}_{ph}^{i}%
}{\partial n^{k}}+\frac{\partial \overset{\left( 3/2\right) }{x}_{ph}^{i}}{%
\partial n^{k}}\right] dn^{k}=0.  \label{v-18}
\end{eqnarray}
All the coefficients are taken at the emission time $t_{e}$. Since both $%
n^{i}$ and $n^{i}+dn^{i}$ are normalized, one has also 
\begin{equation}
n^{k}dn^{k}=0.  \label{v-19}
\end{equation}
Solving the four equations (\ref{v-18}), (\ref{v-19}) gives the four
quantities $dt/dt_{e}$ and $dn^{k}/dt_{e}$. In the time transfer problem we are dealing with, only $%
dt/dt_{e}$ is required.

Making the scalar product of (\ref{v-18}) by $n^{i}$, it turns out that,
using (\ref{v-19}) 
\begin{eqnarray}
\left[ c+n^{i}\left( -v_{B}^{i}-ta_{B}^{i}+\frac{\partial \overset{\left(
1\right) }{x}_{ph}^{i}}{\partial t}\right) \right] dt \notag \\
+n^{i}\left[ v_{A}^{i}-v_{B}^{i}-ta_{B}^{i}-\frac{1}{2}%
t^{2}b_{B}^{i}+v_{A}^{k}\frac{\partial \overset{\left( 1\right) }{x}_{ph}^{i}%
}{\partial x_{A}^{k}}\right] dt_{e}  \notag \\
+n^{i}\left[ \frac{\partial \overset{\left( 1\right) }{x}_{ph}^{i}}{\partial
n^{k}}+\frac{\partial \overset{\left( 3/2\right) }{x}_{ph}^{i}}{\partial
n^{k}}\right] dn^{k}=0    \label{v-20} 
\end{eqnarray}
where the third order terms in the coefficient of $dt$\ have been discarded
since there is no zeroth order term in the coefficients of both $dt_{e}$\
and $dn^{k}$. In (\ref{v-20}), $dn^{k}$ has to be known up to the $c^{-1}$
order only, since there is no zeroth nor first order term in $c^{-1}$ in its
coefficient. This is given by equation (\ref{v-18}), written at the relevant order
\begin{equation}
\left[ cn^{i}-v_{B}^{i}\right] dt+\left[ v_{A}^{i}-v_{B}^{i}\right]
dt_{e}+ctdn^{i}=0.  \label{v-21}
\end{equation}
Inserting (\ref{v-21}) into (\ref{v-20}) to eliminate $dn^{k}$ leads to an
equation which can easily be solved in $dt/dt_{e}$. It turns out that $%
\overset{\left( 3/2\right) }{x}_{ph}^{i}$ leads to terms of order $c^{-2}$
or more. Hence, this term is not needed. Finally
\begin{equation}
\frac{dt}{dt_{e}}=\overset{\left( 1/2\right) }{\frac{dt}{dt_{e}}}+\overset{%
\left( 1\right) }{\frac{dt}{dt_{e}}}+\overset{\left( 3/2\right) }{\frac{dt}{%
dt_{e}}}  \label{v-22}
\end{equation}
with 
\begin{equation}
\overset{\left( 1/2\right) }{\frac{dt}{dt_{e}}}=n^{i}\frac{%
v_{B}^{i}-v_{A}^{i}}{c}  \label{v-23}
\end{equation}
(classical Doppler effect) 
\begin{equation}
\overset{\left( 1\right) }{\frac{dt}{dt_{e}}}=n^{i}\frac{v_{B}^{i}-v_{A}^{i}%
}{c}n^{k}\frac{v_{B}^{k}}{c}+n^{i}\frac{ta_{B}^{i}}{c}  \label{v-24}
\end{equation}
\begin{eqnarray}
&&\overset{\left( 3/2\right) }{\frac{dt}{dt_{e}}} =n^{i}\frac{%
v_{B}^{i}-v_{A}^{i}}{c}\left( n^{k}\frac{v_{B}^{k}}{c}\right) ^{2}+n^{i}%
\frac{2v_{B}^{i}-v_{A}^{i}}{c}n^{k}\frac{ta_{B}^{k}}{c} \notag \\
&&+\frac{1}{2}n^{i}%
\frac{t^{2}b_{B}^{i}}{c}  \label{v-25} -n^{i}\frac{v_{B}^{i}-v_{A}^{i}}{c}\frac{n^{k}}{c}\frac{\partial \overset{%
\left( 1\right) }{x}_{ph}^{k}}{\partial t}-n^{i}\frac{v_{A}^{k}}{c}\frac{%
\partial \overset{\left( 1\right) }{x}_{ph}^{i}}{\partial x_{A}^{k}}  \notag
\\
&&+\left[ -\frac{n^{i}}{ct}\frac{v_{B}^{k}-v_{A}^{k}}{c}+\frac{n^{i}n^{k}}{ct}%
n^{l}\frac{v_{B}^{l}-v_{A}^{l}}{c}\right] \frac{\partial \overset{\left(
1\right) }{x}_{ph}^{i}}{\partial n^{k}}.  
\end{eqnarray}
This expression has now to be inserted in (\ref{v-14}). However, $v_{B}^{2}$
and $W_{B}$ are considered at the reception time in (\ref{v-14}) (where $dt_A=dt_e$), while all
the coefficients in (\ref{v-23})-(\ref{v-25}) are considered at the emission
time. It turns out
\begin{equation}
v_{B}^{2}\left( t_{e}+t\right) =v_{B}^{2}+2tv_{B}^{i}a_{B}^{i} +\mordre{2} \label{v-26}
\end{equation}
\begin{equation}
W_{B}\left( t_{e}+t\right) =W_{B}+tv_{B}^{i}\partial _{i}W_{B}  +\mordre{2} \label{v-27}
\end{equation}
all the terms being considered at time $t_{e}$\ when not precised. One obtains
finally (suppressing the zero indexes)
\begin{eqnarray}
&&\frac{d\tau _{B}}{d\tau _{A}} =1+n^{i}\frac{v_{B}^{i}-v_{A}^{i}}{c}
\label{v-28} \\
&&+\frac{v_{A}^{2}-v_{B}^{2}}{2c^{2}}+\frac{W_{A}-W_{B}}{c^{2}}+n^{i}\frac{%
v_{B}^{i}-v_{A}^{i}}{c}n^{k}\frac{v_{B}^{k}}{c}+n^{i}\frac{cta_{B}^{i}}{c^2} 
\notag \\
&&+n^{i}\frac{v_{B}^{i}-v_{A}^{i}}{c}\left( \frac{v_{A}^{2}-v_{B}^{2}}{2c^{2}%
}+\frac{W_{A}-W_{B}}{c^{2}}\right) \notag \\ &&-tv_{B}^{i}\frac{a_{B}^{i}+ \partial
_{i}W_{B}}{c^{2}}+ \overset{\left( 3/2\right) }{\frac{dt}{dt_{e}}} + \mordre{4} \notag
\end{eqnarray}
where the $dt/dt_{e}$ terms of orders $\left( 1/2\right) $ and $\left(
1\right) $\ are written explicitely to exhibit first and second order
terms, corresponding to classical Doppler effect, special relativistic
contributions and Einstein's gravitational effects. All the $c^{-3}$ terms have been grouped on the third and fourth line of equation (\ref{v-28}). Remark
that $\partial _{i}W_{B}$\ in (\ref{v-27}) and (\ref{v-28}) corresponds to $a_{B}^{i}$ in
the case of free fall motion (i.e. if B\ is a satellite).

\subsubsection{Useful formulae}

Let us consider measurements made from a ground-based station (G), emitting
a photon towards a satellite (S).\ This photon is reflected at (S) and come
back to (G).\ To each event (emission, reflexion, return) is associated a
proper time measured on the involved clock.\ Let $%
\tau _{G}^{-}$ be the proper time on (G) at emission, $\tau _{S}$ the proper
time on (S) at reflexion, and $\tau _{G}^{+}$ the proper time on (G) when
the photon comes back. Two problems may be tackled :

- the ''distance problem'', involving $\tau _{G}^{+}-\tau
_{G}^{-}$ ;

- the ''offset problem'', involving $\left( \tau _{G}^{+}+\tau
_{G}^{-}\right) /2-\tau _{S}$.



Obviously, the (gravitational) theory doesn't give the values of this last
quantity, but allows to calculate $d\tau _{G}^{+}$ and $d\tau
_{S}$ as soon as $d\tau _{G}^{-}$ and the orbits of the ground-based and the onboard clocks are
known. Reciprocally, the measurements of
these quantities lead to constraints on both the orbits and the real
(unperfect) clocks used in the time-measurement process.

Hence, let us define the ''distance problem'' and ''offset problem'' quantities (observables) by, respectively 
\begin{eqnarray*}
DP &=&\frac{d}{d\tau _{G}^{-}}\left( \tau _{G}^{+}-\tau _{G}^{-}\right) \\
OP &=&\frac{d}{d\tau _{G}^{-}}\left( \frac{\tau _{G}^{+}+\tau _{G}^{-}}{2}%
-\tau _{S}\right).
\end{eqnarray*}
For exhaustivity, let us point out the ratio of the ground emitted $\nu
_{G}^{-}$ and the ground received $\nu _{G}^{+}$ frequencies is simply
related to $DP$ by 
\begin{equation*}
\frac{\nu _{G}^{-}}{\nu _{G}^{+}}=\frac{d\tau _{G}^{+}}{d\tau _{G}^{-}}=1+DP.
\end{equation*}

Remark that since both (G) and (S) data are available, these quantities can
be evaluated as functions of both $\tau _{G}^{-}$, $\tau _{G}^{+}$ or $\tau _{S}$. However, in relevant applications, the (G) clock has a precision sensitively better
than the (S) one. Hence, it is well-suited using the (G) clock as reference
time.

\subsubsection{Explicit expressions of $DP$ and $OP$}

Let us calculate $DP$ and $OP$ as functions of the involved
quantities at the emission ($\epsilon =-$) or reception ($\epsilon =+$) time 
$t_{G}^{\epsilon }$ : 
\begin{eqnarray}
DP\left( t_{G}^{\epsilon }\right) &=&\frac{d\tau _{G}^{+}}{d\tau _{G}^{-}}%
\left( t_{G}^{\epsilon }\right) -1 \\
OP\left( t_{G}^{\epsilon }\right) &=&\frac{1}{2}+\frac{1}{2}\frac{d\tau
_{G}^{+}}{d\tau _{G}^{-}}\left( t_{G}^{\epsilon }\right) -\frac{d\tau _{S}}{%
d\tau _{G}^{-}}\left( t_{G}^{\epsilon }\right) \label{eq:offsetproblem}.
\end{eqnarray}
In the relevant applications, the ground based station is the Earth.\ It is
legitimate to consider that, during the time transfer, (1) the Earth is
rigid and uniformly rotating, (2) the Earth gravitational field only is
acting (see discussion in section \ref{sec:model}). Hence, \noindent one has, from (\ref{v-13}) 
\begin{equation*}
\frac{d\tau _{G}^{+}}{d\tau _{G}^{-}}=\frac{dt_{G}^{+}}{dt_{G}^{-}}+O\left(
c^{-4}\right) .
\end{equation*}
This shows the shift between the emitted and coming back
frequencies of the photon doesn't require an
Earth model (for both potential and rotation). The interpretation is obvious, but this doesn't explicitely appear
in formulae available in the Blanchet et al paper \cite{bstw01}.\\
Let us put 
\begin{eqnarray}
X^{i} &=&x_{G}^{i}-x_{S}^{i}  \label{v-32} \\
V^{i} &=&v_{G}^{i}-v_{S}^{i}  \notag \\
A^{i} &=&a_{G}^{i}-a_{S}^{i}  \notag \\
B^{i} &=&b_{G}^{i}-b_{S}^{i}  \notag \\
D &=&\sqrt{X^{i}X^{i}}  \notag \\
N^{i} &=&X^{i}/D  \notag \\
P^{ik} &=&\delta ^{ik}-N^{i}N^{k}  \notag
\end{eqnarray}
One finds, for the involved quantities 
\begin{eqnarray*}
\frac{d\tau _{G}^{+}}{d\tau _{G}^{-}}\left( t_{G}^{\epsilon }\right)
&=&1+c^{-1}\overset{\left( 1/2\right) }{D}^{\epsilon }+c^{-2}\overset{\left(
1\right) }{D}^{\epsilon }+c^{-3}\overset{\left( 3/2\right) }{D}^{\epsilon }
\\
\frac{d\tau _{S}}{d\tau _{G}^{-}}\left( t_{G}^{\epsilon }\right) &=&1+c^{-1}%
\overset{\left( 1/2\right) }{E}^{\epsilon }+c^{-2}\overset{\left( 1\right) }{%
E}^{\epsilon }+c^{-3}\overset{\left( 3/2\right) }{E}^{\epsilon }
\end{eqnarray*}
with 
\begin{eqnarray*}
\overset{\left( 1/2\right) }{D}^{\epsilon } &=&2N^{i}V^{i} \\
\overset{\left( 1\right) }{D}^{\epsilon } &=&2\left( 1+\epsilon \right)
\left( N^{i}V^{i}\right) ^{2}-2\epsilon V^{i}V^{i}-2\epsilon DN^{i}A^{i} \\
\overset{\left( 3/2\right) }{D}^{\epsilon } &=&N^{i}V^{i}\left[
v_{G}^{k}v_{G}^{k}+\left( N^{k}v_{G}^{k}\right) ^{2}\right]\\
&&+2N^{i}v_{S}^{i}P^{kl}v_{G}^{k}V^{l}-2\left( N^{i}V^{i}\right)
^{2}N^{k}v_{S}^{k} \\
&&+D\left[
2v_{G}^{k}a_{G}^{k}+2N^{k}v_{G}^{k}N^{l}a_{G}^{l}+P^{ik}V^{i}a_{G}^{k}+3V^{i}A^{i}%
\right] \\
&&-\left[ \left( 3+4\epsilon \right) N^{i}V^{i}+2N^{i}v_{G}^{i}\right] \left[
DN^{i}A^{i}+P^{ik}V^{i}V^{k}\right] \\
&&+D^{2}N^{i}\left[ b_{G}^{i}+B^{i}\right] \\
&&-\frac{2}{D}P^{ik}V^{i}c^{2}\overset{\left( 1\right) }{x}%
_{ph}^{k}-2N^{i}V^{i}N^{k}\frac{c^{2}\partial \overset{\left( 1\right) }{x}%
_{ph}^{k}}{c\partial t} \\
&&-\frac{2}{D}N^{k}P^{il}V^{i}\frac{c^{2}\partial \overset{\left( 1\right) }{%
x}_{ph}^{k}}{\partial n^{l}}-2N^{k}v_{S}^{l}\frac{c^{2}\partial \overset{%
\left( 1\right) }{x}_{ph}^{k}}{\partial x_{S}^{l}}
\end{eqnarray*}
and 
\begin{eqnarray*}
\overset{\left( 1/2\right) }{E}^{\epsilon } &=&N^{i}V^{i} \\
\overset{\left( 1\right) }{E}^{\epsilon } &=&-\left( \frac{1}{2}+\epsilon
\right) V^{i}V^{i}+\left( 1+\epsilon \right) \left[ \left( N^{i}V^{i}\right)
^{2}-DN^{i}a_{G}^{i}\right] \\
&&+\epsilon DN^{i}a_{S}^{i}+W_{G}-W_{S} \\
\overset{\left( 3/2\right) }{E}^{\epsilon } &=&N^{i}v_{S}^{i}V^{k}v_{S}^{k}+%
\frac{1}{2}N^{i}V^{i}\left[ v_{G}^{k}v_{G}^{k}-\left( N^{k}v_{S}^{k}\right)
^{2}\right] \\
&&-2\left( 1+\epsilon \right) N^{i}V^{i}P^{kl}V^{k}V^{l} \\
&&+D\left[ -\frac{1}{2}V^{i}a_{S}^{i}+\left( 1+\epsilon \right) V^{i}\left(
2a_{G}^{i}-a_{S}^{i}\right) \right] \\
&&+D\left[ -\frac{1}{2}%
N^{i}V^{i}N^{k}a_{S}^{k}+N^{i}v_{S}^{i}N^{k}a_{S}^{k}\right]\\
&&-2\left( 1+\epsilon
\right) D N^{i}V^{i}N^{k}A^{k} \\
&&+D^{2}\left[ -\frac{1}{2}N^{i}b_{S}^{i}+\left( 1+\epsilon \right)
N^{i}b_{G}^{i}\right] \\
&&+N^{i}V^{i}\left( W_{G}-W_{S}\right) +\epsilon Dv_{S}^{i}\partial _{i}W_{S}
\\
&&-\frac{1}{D}P^{ik}V^{i}c^{2}\overset{\left( 1\right) }{x}%
_{ph}^{k}-N^{i}V^{i}N^{k}\frac{c^{2}\partial \overset{\left( 1\right) }{x}%
_{ph}^{k}}{c\partial t} \\
&&-\frac{1}{D}N^{k}P^{il}V^{i}\frac{c^{2}\partial \overset{\left( 1\right) }{%
x}_{ph}^{k}}{\partial n^{l}}-N^{k}v_{S}^{l}\frac{c^{2}\partial \overset{%
\left( 1\right) }{x}_{ph}^{k}}{\partial x_{S}^{l}}
\end{eqnarray*}
The quantities $D,N,v_{G},v_{S},a_{G},a_{S},b_{G},b_{S}$ are taken at
the time $t_{G}^{\epsilon }$, while the photon first order term $\overset{%
\left( 1\right) }{x}_{ph}$ and its derivatives have to be computed for $%
t=+D/c$, $n=+\Delta x/D$ and $x_{S}$ (at time $t_{G}^{\epsilon }$) for both $%
\epsilon =\pm $. Let us recall that $c^{2}\overset{\left( 1\right) }{x}%
_{ph}$ and $ct$ are zeroth order quantities.\\
From these expressions, it turns out that 
\begin{equation*}
OP\left( t_{G}^{\epsilon }\right) =c^{-2}\overset{\left( 1\right) }{OP}%
^{\epsilon }+c^{-3}\overset{\left( 3/2\right) }{OP}^{\epsilon }
\end{equation*}
with 
\begin{eqnarray*}
\overset{\left( 1\right) }{OP}^{\epsilon } &=&\frac{1}{2}%
V^{i}V^{i}+DN^{i}a_{G}^{i}-W_{G}+W_{S} \\
\overset{\left( 3/2\right) }{OP}^{\epsilon } &=&-N^{i}V^{i}\left(
N^{k}v_{S}^{k}\right) ^{2}-\frac{1}{2}V^{i}V^{i}N^{k}V^{k}+\left(
N^{i}V^{i}\right) ^{3} \\
&&+D\left[ v_{G}^{i}a_{G}^{i}-\epsilon V^{i}\left( A^{i}+a_{G}^{i}\right) +%
\frac{1}{2}N^{i}V^{i}N^{k}a_{S}^{k}\right] \\
&&-\epsilon D^{2}N^{i}b_{G}^{i} \\
&&-N^{i}V^{i}\left( W_{G}-W_{S}\right) -\epsilon Dv_{S}^{i}\partial
_{i}W_{S}.
\end{eqnarray*}


\section{Conclusion and discussion}

In this paper, one has proposed formulae involved in two way time transfer
problems in a form directly adapted for practical uses.\ All these
quantities are expressed as functions of quantities defined at a unique
(emission or reception) ground-based time. Considering foreseeable increase
of clock's accuracy, the formulae are given including the $J_{2}$
contribution, but can be restricted to the spherical case making $J_{2}=0$,
which is sufficient for the time being applications. 

In this paper, the chosen unique time is a ground
time because ground based clocks have generally precisions sensitively better than on
board clocks.\ However, in the case where one needs to make some time
transfer between two different ground based stations, distant in such a way
there is no spacecraft in common view, the link will involve an on board
clock, with the resulting limitation in precision. Let $%
t_{1}$ and $t_{2}$\ be the times of the links between the spacecraft and
the first and second ground stations.\ Since the spacecraft motion is
considered during the time interval $\left[ t_{1},t_{2}\right] $, the
crossed $cdtdx^{i}$ term in (\ref{iv-1}) is a $c^{-4}$ order term. Hence, the onboard proper time formula reduces
to
\begin{equation*}
\left( \frac{d\tau }{dt}\right) ^{2}=1+\frac{2w}{c^{2}}-\frac{v^{2}}{c^{2}}%
+\mordre{4}
\end{equation*}
leading to the spacecraft proper time interval
\begin{equation*}
\Delta \tau \left( t_{1},t_{2}\right) =t_{2}-t_{1}+\frac{1}{c^{2}}%
\int_{t_{1}}^{t_{2}}\left( w-\frac{v^{2}}{2}\right) dt+\mordre{4} .
\end{equation*}
Let us enforce the fact that, since the time interval $\left[ t_{1},t_{2}%
\right] $\ is finite, the orbit used in the previous r.h.s. member integral
cannot be developped around the emission time, and should even include
all the perturbations acting on the satellite's motion.\ On the other hand,
a classical description of this orbit is sufficient, since relativistic
corrections generate $c^{-4}$ contributions to $\Delta \tau$.

\begin{acknowledgments}
Olivier Minazzoli wants to thank the Government of the Principality of Monaco for their financial support.
\end{acknowledgments}

\end{document}